\documentclass[11pt]{article}
\usepackage{times}
\usepackage{sidecap}
\usepackage{geometry}
\usepackage{floatrow}
\usepackage{float}
\usepackage{hyperref}
\usepackage{graphics,graphicx}
\usepackage{subcaption}
\usepackage{epsf,wrapfig}
\usepackage{natbib,bm}
\usepackage[misc]{ifsym}
\usepackage{bbding}
\usepackage[T1]{fontenc}

\usepackage{paralist}
\let\olditem\item
\renewenvironment{thebibliography}[1]{%
  \section*{\refname}
  \let\par\relax
  \renewcommand{\item}[1][]{\olditem[\textbullet]}%
  \inparaenum}{\endinparaenum}

\newcounter{aff}

\DeclareGraphicsExtensions{.jpg,.pdf,.png,.eps}

\usepackage{caption}
\usepackage[dvipsnames]{xcolor}
\usepackage{tcolorbox}
\bibpunct{(}{)}{;}{a}{}{,}
\usepackage{soul}
\usepackage[version=4]{mhchem}

\definecolor{darkcyan}{rgb}{0.0, 0.0, 0.7}

\definecolor{boxcolor}{RGB}{193, 229, 255}
\definecolor{figboxcolor}{RGB}{193, 229, 255}
\definecolor{boxcolorgray}{RGB}{227, 227, 227}
\definecolor{darkcyan}{rgb}{0.0, 0.0, 0.7}

\sethlcolor{boxcolorgray}

\def\gtrsim{\lower 2pt \hbox{$\, \buildrel {\scriptstyle >}\over
{\scriptstyle \sim}\,$}}
\def\lesssim{\lower 2pt \hbox{$\, \buildrel {\scriptstyle <}\over
{\scriptstyle \sim}\,$}}

\def\approxlt{\lower.2em\hbox{$\buildrel < \over \sim$}}
\def\approxgt{\lower.2em\hbox{$\buildrel > \over \sim$}}

\def\arcsec{$^{\prime\prime}$}

\def\be{\begin{equation}}
\def\ee{\end{equation}}
\def\bea{\begin{eqnarray*}}
\def\eea{\end{eqnarray*}}

\def\arcsec{^{\prime\prime}}

\def\jref@jnl#1{{\rm#1}}
\def\aj{\jref@jnl{AJ}}			
\def\araa{\jref@jnl{ARA\&A}}		
\def\apj{\jref@jnl{ApJ}}			
\def\apjl{\jref@jnl{ApJ}}		
\def\apjs{\jref@jnl{ApJS}}		
\def\ao{\jref@jnl{Appl.~Opt.}}		
\def\apss{\jref@jnl{Ap\&SS}}		
\def\aap{\jref@jnl{A\&A}}		
\def\aapr{\jref@jnl{A\&A~Rev.}}		
\def\aaps{\jref@jnl{A\&AS}}		
\def\azh{\jref@jnl{AZh}}			
\def\baas{\jref@jnl{BAAS}}		
\def\jrasc{\jref@jnl{JRASC}}		
\def\memras{\jref@jnl{MmRAS}}		
\def\mnras{\jref@jnl{MNRAS}}		
\def\pra{\jref@jnl{Phys.~Rev.~A}}	
\def\prb{\jref@jnl{Phys.~Rev.~B}}	
\def\prc{\jref@jnl{Phys.~Rev.~C}}	
\def\prd{\jref@jnl{Phys.~Rev.~D}}	
\def\pre{\jref@jnl{Phys.~Rev.~E}}	
\def\prl{\jref@jnl{Phys.~Rev.~Lett.}}	
\def\pasp{\jref@jnl{PASP}}		
\def\pasj{\jref@jnl{PASJ}}		
\def\qjras{\jref@jnl{QJRAS}}		
\def\skytel{\jref@jnl{S\&T}}		
\def\solphys{\jref@jnl{Sol.~Phys.}}	
\def\sovast{\jref@jnl{Soviet~Ast.}}	
\def\ssr{\jref@jnl{Space~Sci.~Rev.}}	
\def\zap{\jref@jnl{ZAp}}			
\def\nat{\jref@jnl{Nature}}		
\def\iaucirc{\jref@jnl{IAU~Circ.}}
\def\aplett{\jref@jnl{Astrophys.~Lett.}}
\def\apspr{\jref@jnl{Astrophys.~Space~Phys.~Res.}}
\def\bain{\jref@jnl{Bull.~Astron.~Inst.~Netherlands}}
\def\fcp{\jref@jnl{Fund.~Cosmic~Phys.}}
\def\gca{\jref@jnl{Geochim.~Cosmochim.~Acta}}
\def\grl{\jref@jnl{Geophys.~Res.~Lett.}}
\def\jcp{\jref@jnl{J.~Chem.~Phys.}}	
\def\jgr{\jref@jnl{J.~Geophys.~Res.}}	
\def\jqsrt{\jref@jnl{J.~Quant.~Spec.~Radiat.~Transf.}}
\def\memsai{\jref@jnl{Mem.~Soc.~Astron.~Italiana}}
\def\nphysa{\jref@jnl{Nucl.~Phys.~A}}
\def\physrep{\jref@jnl{Phys.~Rep.}}
\def\physscr{\jref@jnl{Phys.~Scr}}
\def\planss{\jref@jnl{Planet.~Space~Sci.}}
\def\procspie{\jref@jnl{Proc.~SPIE}}

\def\arcsec{\hbox{$^{\prime\prime}$}}

\def\ion#1#2{#1$\;${\small\rm\@Roman{#2}}\relax}

\newcount\lecurrentfam
\def\LaTeX{\lecurrentfam=\the\fam \leavevmode L\raise.42ex
\hbox{$\fam\lecurrentfam\scriptstyle\kern-.3em A$}\kern-.15em\TeX}
\def\sizrpt{
(\fontname\the\font): em=\the\fontdimen6\font, ex=\the\fontdimen5\font
\typeout{
(\fontname\the\font): em=\the\fontdimen6\font, ex=\the\fontdimen5\font
}}
\definecolor{dark}{rgb}{ 0.34, 0.09, 0.49}

\usepackage{fancyhdr,url}
\usepackage{transparent}
\usepackage{tablefootnote}
\usepackage{mathpazo}

\begin{document}

\noindent
\begin{raggedright}
{\Huge\bfseries Multi-wavelength Polarization as a Probe of Planet Formation 
}\end{raggedright}\\

\medskip
\noindent
Ian Stephens$^{1,}$\textsuperscript{\Letter}, Sarah Sadavoy$^{2,}$\textsuperscript{\Letter}, Akimasa Kataoka$^{3,}$\textsuperscript{\Letter}, Zhe-Yu Daniel Lin$^4$, Boy Lankhaar$^5$, Greta Guidi$^6$, Takahiro Ueda$^3$, Josep Miquel Girart$^7$, Gary Fuller$^{8,9}$, Ian Czekala$^{10}$, Enrique Macias$^{11}$, Claudia Toci$^{12}$, Kiyoaki Doi$^{13}$, Korash Assani$^{14}$

\vspace{1em}
{
\footnotesize{
\noindent $^{1}$ Department of Earth, Environment, and Physics, Worcester State University, USA; 
$^{2}$ Department of Physics, Engineering and Astronomy, Queen's University, Canada; 
$^{3}$ National Astronomical Observatory of Japan, Japan; 
$^{4}$ National Radio Astronomy Observatory, USA;
$^{5}$ Institute of Theoretical Astrophysics, University of Oslo, Norway;
$^{6}$ Univ.~Grenoble Alpes, CNRS, Institut de Planetologie et d'Astrophysique de Grenoble, France; 
$^{7}$ Institut de Ci\`encies de l'Espai, CSIC, IEEC, Catalonia, Spain; 
$^{8}$ Jodrell Bank Centre for Astrophysics, Department of Physics;
\& Astronomy, The University of Manchester, UK;
$^{9}$ Department of Physics, University of Cologne, Germany; 
$^{10}$ Australia Telescope National Facility, CSIRO Space and Astronomy, Australia; 
$^{11}$ European Southern Observatory, Germany; 
$^{12}$ Departamento de Fisica Aplicada III, ETSI, Universidad de Sevilla, Spain; 
$^{13}$ Max Planck Institute for Astronomy, Germany; 
$^{14}$ DCBP, Universität Bern, Switzerland \\

\noindent \Letter~\url{istephens@worcester.edu}; 
\Letter~\url{sarah.sadavoy@queensu.ca};
\Letter~\url{akimasa.kataoka@nao.ac.jp};

}
}
\vspace{-3mm}
\begin{tcolorbox}[boxsep=3pt,left=0pt,right=0pt,top=1pt,bottom=1pt,colframe=boxcolor,colback=boxcolor]
\justify
\textbf{Abstract:} 
Polarization reveals details of circumstellar disks that are not readily available from total intensity observations. Via multi-wavelength polarimetric observations of disks, we can trace how planets form, how dust grains evolve at the onset of planet formation, and how magnetic fields (B-fields) influence accretion. Since polarized emission from these disks is typically at levels from 0.5\% to a few percent of the total emission, very high sensitivities are needed. Current ALMA capabilities make these key measurements of protoplanetary disks challenging due to (1) long integrations with separate observations to measure multi-wavelength polarization of substructures, (2) prohibitive time requests to observe large samples of disks, (3) limited sensitivity and frequency coverage for polarization in spectral lines; and (4) insufficient calibration of polarization in all the Stokes parameters. An ALMA2040 with large collecting area, increased bandwidth with multi-band capabilities, and accurate polarization calibration will provide a great leap in our understanding of the formation of protoplanets through polarization. Specifically, such an ALMA2040 can use polarized emission from the continuum and spectral lines to: (A) provide stringent tests of streaming instabilities; (B) accurately constrain grain shape, size, porosity, and alignment; (C) resolve polarization of substructures for a large sample of disks with resolutions less than 5\,au; (D) quickly survey 100s of disks; and (E) measure the B-field strength and morphology of disks.

\end{tcolorbox}

\newpage{}

\section{Scientific context and open questions}
Polarization observations provide a powerful view of protoplanetary disks that are more informative than total-intensity measurements alone. At every spatial resolution element, polarimetric observations measure not only the total intensity (Stokes $I$), but also the linear polarized intensity, the polarization position angle, and the circular polarization, providing up to four independent observational constraints that any physical model of the disk must simultaneously reproduce. As such, polarization observations place substantially stronger constraints on the physical conditions and properties of disks that can shape the formation and evolution of planetary systems. The polarized flux is typically only a small percentage of the total flux ($\sim$1\%), and it changes with both frequency and substructure  in the disks \citep[e.g.,][]{Stephens2017,Stephens2023,Lin2024a}. Given ALMA's current capabilities, we are limited in the disks we can study in polarization by the need for very high sensitivities with both multi-wavelength and high-resolution observations.

With ALMA2040's increased sensitivity and multi-band capabilities, we will answer:

\begin{enumerate}
    \item[A.] What are the physical properties of disks when combining multi-wavelength total intensity with polarization measurements?
    
    \item[B.] Under what conditions does planet formation occur, including the onset and evolution of streaming instabilities?
    
    \item[C.] What are the characteristic dust properties (e.g., size, shape, composition, and alignment) across a broad population of circumstellar disks, and how do they evolve through time?
    
    \item[D.] What is the strength and morphology of magnetic fields (hereafter, B-fields) in protoplanetary disks, and what role do they play in disk evolution and planet formation?
\end{enumerate}

\section{The Present Understanding of Disk Polarization}\label{sec:understanding}

\subsection{Dust Emission}\label{sec:dustemission}
Grains in the interstellar medium are expected to align with their short axes parallel to the B-field via radiative torques (B-RATs; e.g., \citealt{Andersson2015}), making thermal dust emission polarized perpendicular to the B-field. However, B-RAT alignment within circumstellar disks becomes less efficient since grains are bigger and densities are larger, both of which greatly increase the alignment time relative to the gas damping time \citep{Yang2021}. Consistent with this, expected B-RAT polarization patterns  have not been clearly seen in disks. Although there are some cases where polarization may trace dust alignment with B-fields, such detections are tentative and only apply to parts of disks 
\citep[e.g.,][]{Alves2018, Sadavoy2018,Ohashi2025}.

\begin{figure*}[ht!]
\begin{center}
\includegraphics[height=0.215\columnwidth]{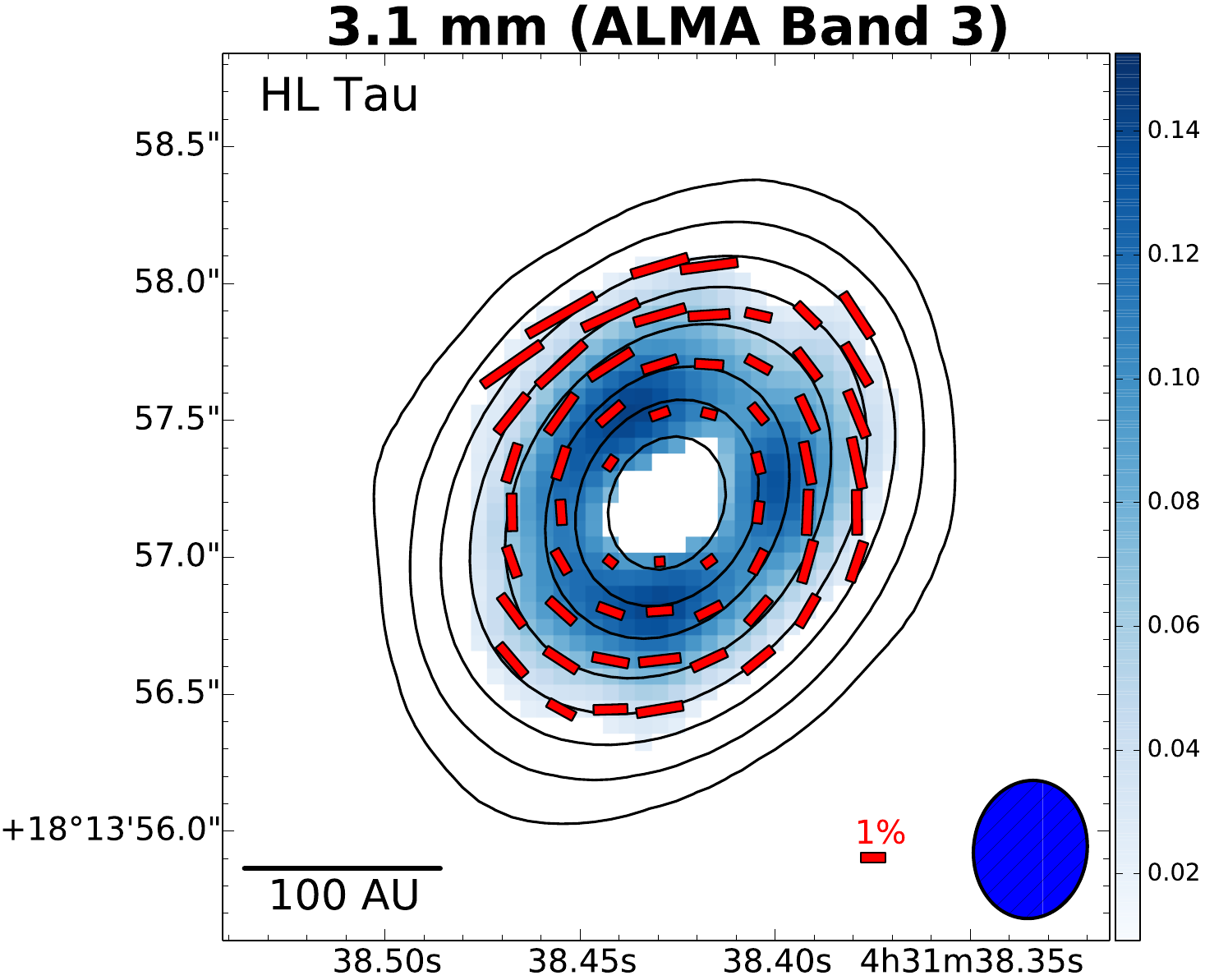}~
\includegraphics[height=0.215\columnwidth]{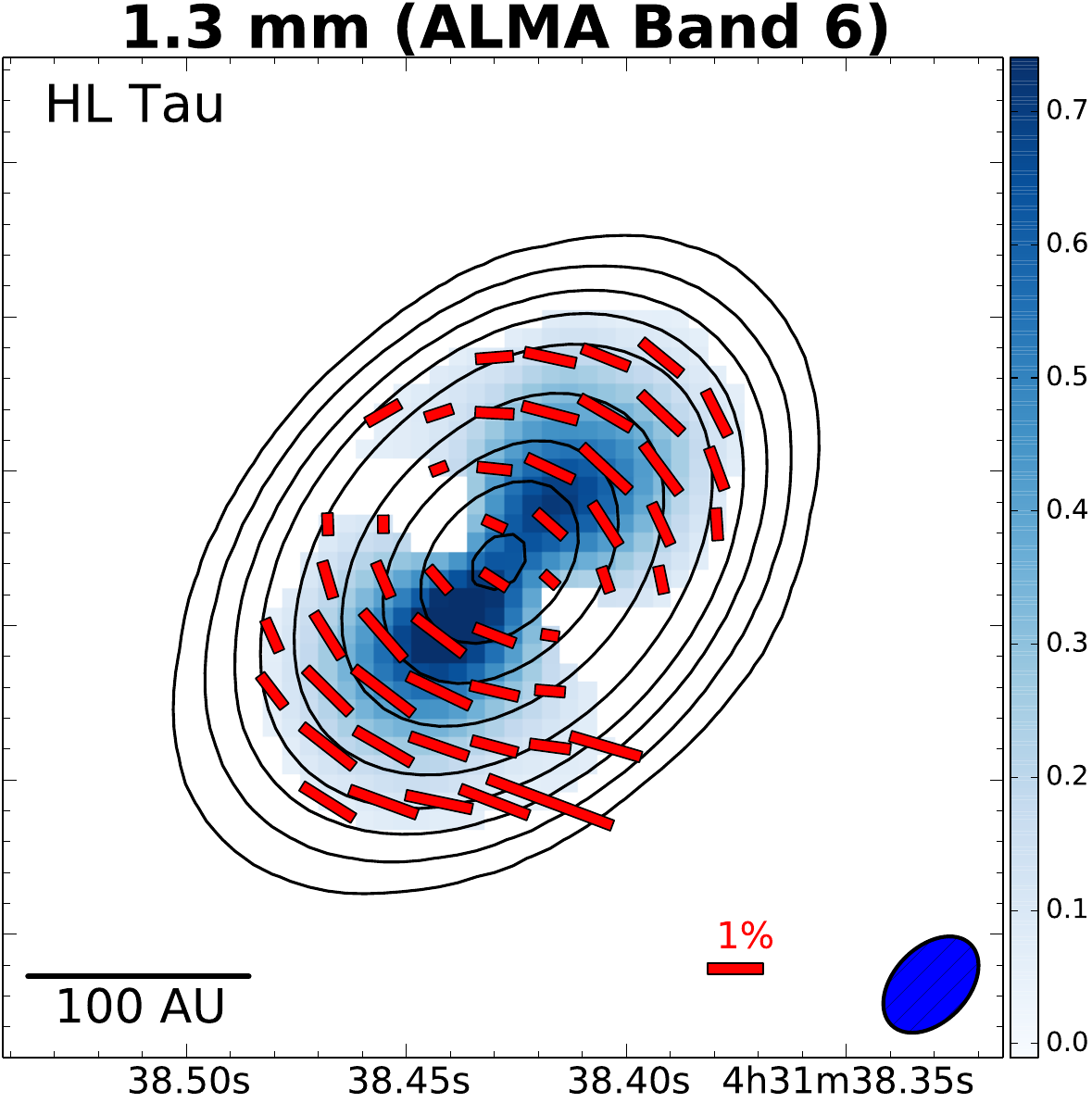}~
\includegraphics[height=0.215\columnwidth]{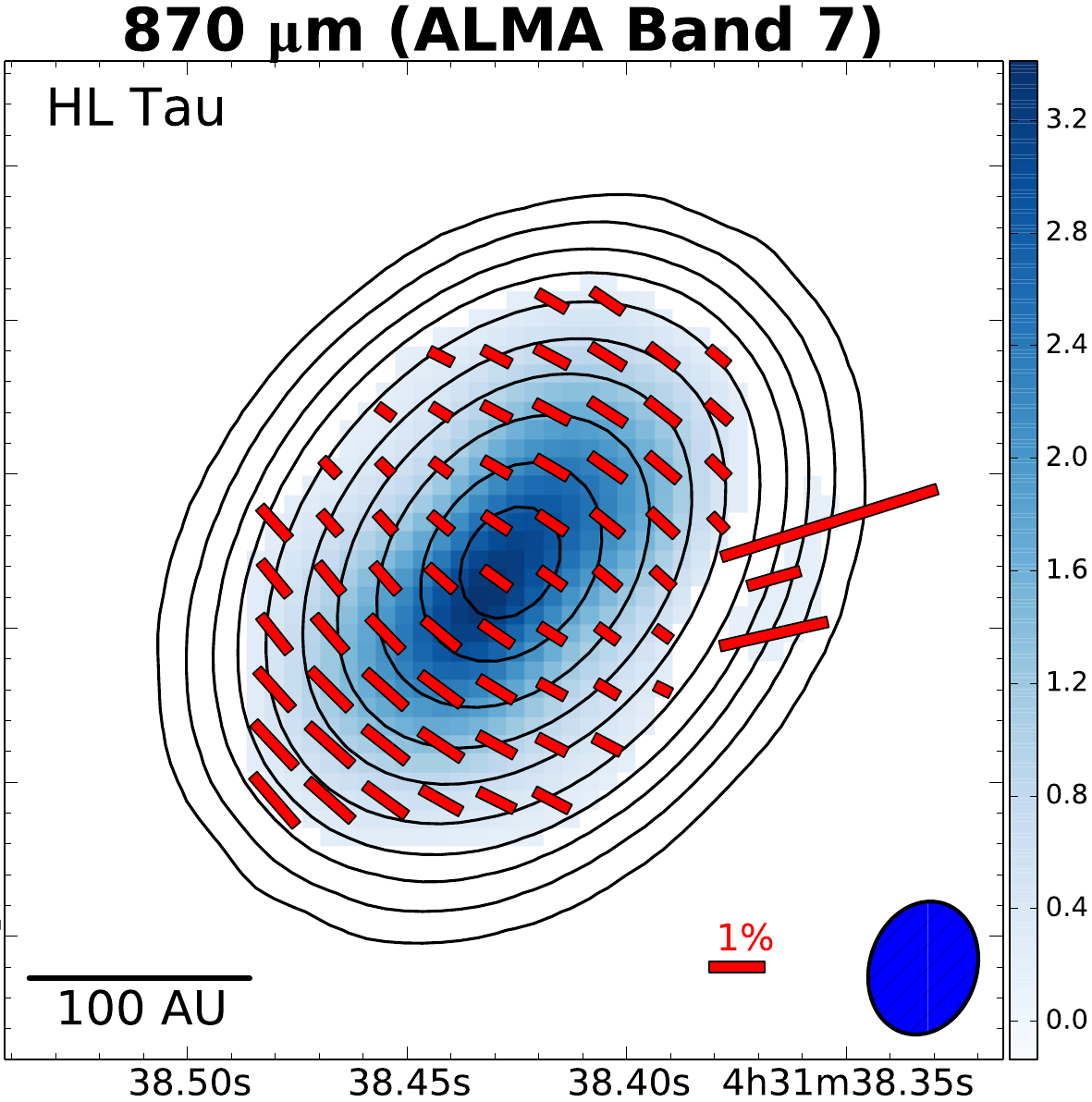}
\includegraphics[height=0.22\columnwidth]{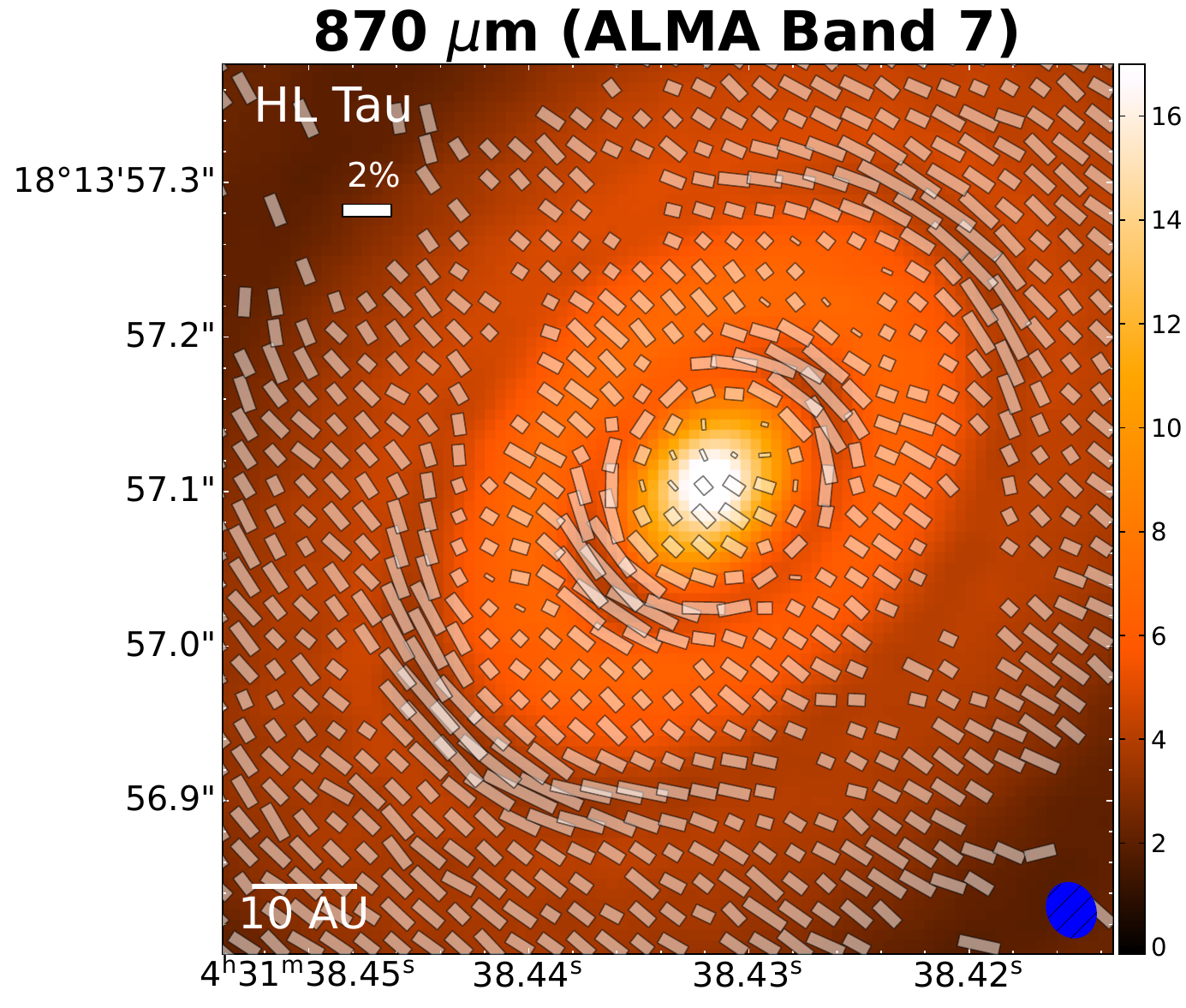}
\end{center}
\vspace{-8pt}
\caption{ALMA polarization observations at 3\,mm ({\bf left}), 1.3\,mm ({\bf middle left}), and 870\,$\mu$m ({\bf middle right}) of HL Tau \citep{Stephens2017}. The contours show Stokes I, color scale shows the polarized intensity (mJy\,bm$^{-1}$), and vectors show the polarization, where the length is proportional to the polarization fraction. {\bf Right:} High resolution zoom-in of HL Tau Band~7 polarization at 34.5 mas (5\,au) resolution, overlaid on a Stokes~I color scale (mJy\,bm$^{-1}$), revealing morphological differences between rings and gaps \citep{Stephens2023}.
}
\label{fig:pol_images} 
\vspace{-12pt}
\end{figure*}

The first three panels of Figure~\ref{fig:pol_images} show the polarization patterns toward HL Tau at ALMA Bands 3, 6, and 7 \citep{Stephens2017}, which represent the general morphologies of these wavelengths seen toward most disks. The uniform polarization along the disk's minor axis seen in Band~7 is expected for polarization by self-scattering for inclined disks \citep[e.g.,][]{Kataoka2015,Yang2016a}.  The azimuthal polarization pattern seen in Band~3 is consistent with grains aligned azimuthally, although the mechanism for this alignment is unclear; the grains may be aligned mechanically/aerodynamically \citep[e.g.,][]{Lin2024b}, by radiation anisotropy \citep[k-RAT;][]{Tazaki2017}, or via grains with wrong internal alignment \citep{Thang2024}. 


Substructure also complicates the interpretation of the polarization. The highest resolution polarization observations to date are toward HL Tau at Band~7 at 34.5 mas (5\,au) resolution (Figure~\ref{fig:pol_images}, right; \citealt{Stephens2023}). The rings of HL Tau show signs of scattering, while the gaps are a mix of scattering plus grain alignment. Beam-smearing of substructure can be difficult to interpret, necessitating sensitivities that allow ALMA to efficiently observe at high resolutions ($\sim$5--20\,au). 


\subsection{Spectral Lines}


In the current paradigm, dust emission does not appear to reliably trace B-fields in disks as it does in the interstellar medium. 
Alternatively, one can investigate the B-field via spectral line polarization.  These methods include the Zeeman effect, the Goldreich-Kylafis effect \citep[GK effect;][]{Goldreich1981}, maser emission \citep{Norris1998}, and ground state alignment \citep[GSA; ][]{Yan2012}.  Detections with these methods have been limited with current ALMA capabilities, but each provides valuable insights into B-fields and the physical conditions of disks.  

The Zeeman effect uniquely provides a direct measurement of the B-field strength.  It requires accurate measurements of the Stokes~$V$ polarization of spectral lines with a high Zeeman-splitting factor. So far, the Zeeman effect of the CN spectral line has remained undetected with ALMA \citep{Vlemmings2019,Harrison2021}, suggesting that B-field strengths in disks are no more than a few mG.  Alternatively, the combination of Zeeman broadening and linear polarization can provide information about the 3-D B-field in a disk \citep{LankhaarTeague2023}. Following this approach, \citet{Teague2025} estimated a B-field strength of about 10~mG in TW Hya. Confirmation of this measurement as well as reliable B-field strengths from line polarization in more disks will require improved sensitivities and calibration of Stokes V.

Similarly, polarization observations of other spectral lines have had limited success with current ALMA capabilities.  The GK effect can be used to trace the B-field morphology similar to linear dust polarization \citep{Girart1999}. The GK effect with CO in disks has had only marginal detections, suggesting polarization levels on the order of 1\% \citep{Stephens2020,Teague2021}. The polarization angles with respect to the B-field due to the GK effect suffer from a 90$^\circ$ ambiguity, requiring statistically meaningful samples in coordination with forward modeling of the disk geometry.
Masers present around disks of high-mass stars \citep{Sanna2010,Moscadelli2011} can also trace B-fields via highly accurate measurements of Stokes~$V$, which is currently not possible with ALMA \citep{Lankhaar2018}.  Finally, disk B-fields can be inferred from GSA of atomic lines \citep[e.g.,][]{Yan2012,Zhang2018}, which has not been attempted in disks as it requires polarization observations for frequencies higher than provided with Band~7.

\section{Science case for the 2040s}
\noindent {\bf A.~Power of simultaneous multi-wavelength polarization observations} -- 
Total-intensity observations alone are insufficient to constrain disk models, as similar continuum spectra can arise from different combinations of grain size, porosity, composition, temperature, and optical depth. Polarization provides a powerful way to break these degeneracies because both the polarization fraction and morphology are sensitive to dust properties and the underlying polarization mechanism. As illustrated in Figure~\ref{fig:pol_images}, the continuum brightness, polarization fraction, and polarization morphology change substantially with wavelength. 
The wavelength dependence of polarization provides a powerful discriminator between self-scattering and grain alignment, while simultaneously constraining grain properties and internal structure. 
At present, obtaining sufficiently deep polarimetric observations in multiple bands requires separate, lengthy observations with substantial overheads. An ALMA2040 able to provide simultaneous multi-band polarization observations would dramatically improve both observing efficiency and the scientific return of every polarization program.\\


\noindent {\bf B.~Identifying sites for planetesimal formation through the Streaming Instability} -- 
%
%
Streaming Instability (SI) is a dynamical process that can circumvent barriers to the growth of planets from micron-sized grains to kilometer-sized planetesimals \citep[e.g.,][]{Drazkowska2023ASPC..534..717D}.
SI results from coupled aerodynamic drag between the gas and dust such that a local concentration of dust can quickly form and then collapse gravitationally \citep{Johansen2007ApJ...662..627J, Johansen2007Natur.448.1022J}. 

Dust polarization provides a unique opportunity to identify SI in protoplanetary disks. 
A grain experiencing a systematic gaseous flow, i.e., the aerodynamic field, can be aligned such that its long axis is parallel to this aerodynamic field \citep{Wurm2000ApJ...529L..57W, Lin2024b}. Polarized radiative-transfer simulations of SI show that SI produces a unique azimuthal aerodynamic field which can explain the observed azimuthally aligned, effectively prolate grains, whereas other dynamical processes fall short \citep{Lin2026}. 

Increased sensitivity and angular resolution will enable direct tests of the alignment theory and empirically probe sites of SI.  
Adequate spatial resolution with multiwavelength polarization avoids beam smearing of polarization between rings and gaps and also permits mapping regions of efficient alignment \citep{Fourkas2026ApJ...999....4F}.
Access to polarization in the inner few-au region will enable studies of SI near the water-ice snowline, where models of gas and dust evolution predict different locations for efficient planetesimal formation \citep{Drazkowska2023ASPC..534..717D}. Polarization interior to the snowline will probe the scattering properties and constrain SI of material relevant to rocky planetesimal formation. Polarization surveys spanning a range of ages, spatial extents, and star-forming regions can determine when and where SI can occur. \\


\begin{figure}[t]
  \centering
  \includegraphics[width=1\textwidth]{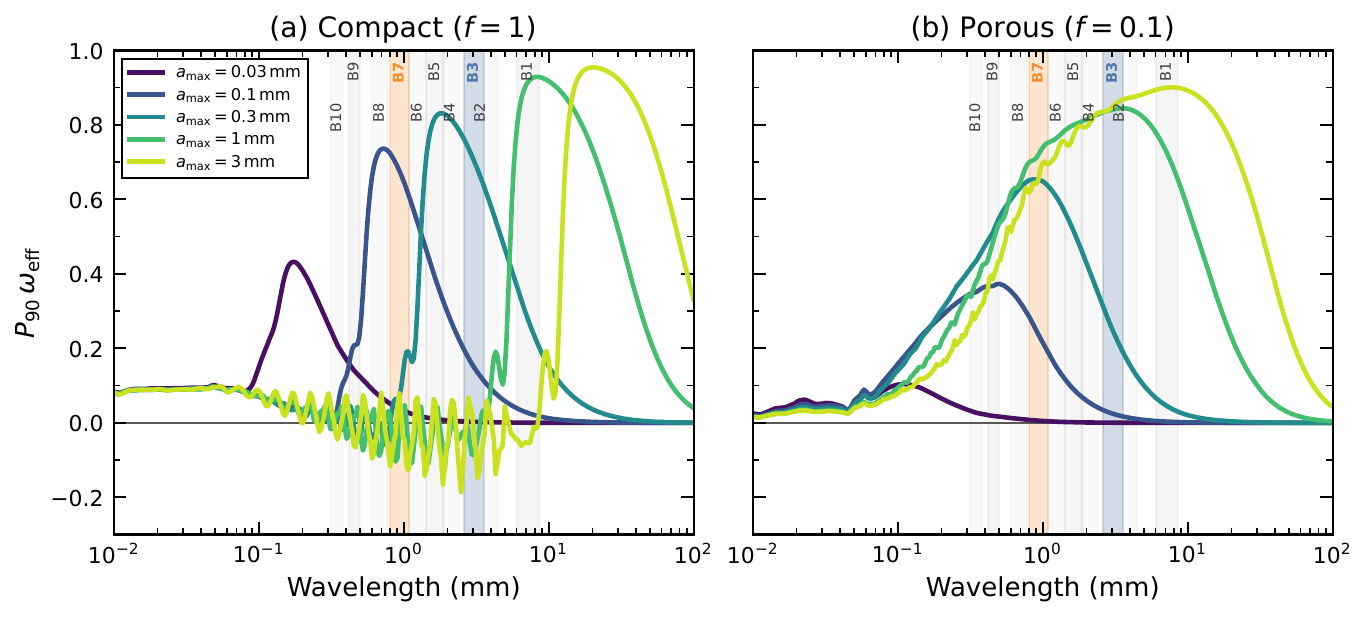}
  \caption{Self-scattering polarization efficiency, $P_{90}\omega_{\rm eff}$, as a function of wavelength for maximum aggregate radii $a_{\max}=0.03$, 0.1, 0.3, 1, and 3\,mm. The left and right panels show compact aggregates with filling factor $f=1$ and porous aggregates with $f=0.1$, respectively, using the standard DSHARP dust composition \citep{Birnstiel2018}; shaded regions mark ALMA Bands~1--10. Here $P_{90}$ is the single-scattering polarization at a scattering angle of $90^\circ$, and $\omega_{\rm eff}$ is the effective albedo. The shift and broadening of the spectral features demonstrate how multi-band polarimetry can break the size--porosity degeneracy. This product is a diagnostic of the dust properties rather than the polarization fraction of a specific disk, which also depends on radiation-field anisotropy, disk geometry, optical depth, and angular resolution.}
  \label{fig:scattering_efficiency}
\end{figure}

\noindent {\bf C.~Evolution of dust properties during planet formation} -- 
%
Dust coagulation is the first step toward planetesimal formation. Whether collisions lead to sticking, bouncing, or fragmentation depends strongly on aggregate composition and porosity. In particular, collision simulations show that the threshold mass for bouncing drops steeply as the filling factor increases, implying that more porous aggregates are less susceptible to bouncing \citep{Oshiro2025}. Measuring both aggregate size and internal structure is therefore essential for identifying viable growth pathways. The millimeter continuum spectral slope alone cannot determine these properties: grain size and porosity are degenerate with the size distribution, composition, temperature, and optical depth.

Dust self-scattering provides a complementary diagnostic. Scattering polarization is most efficient when the size parameter $2\pi a/\lambda$ is of order unity, and its wavelength dependence changes with aggregate porosity \citep{Kataoka2015,Tazaki2019}. Figure~\ref{fig:scattering_efficiency} shows how the wavelength dependence of polarized emission can distinguish grain size from porosity. Compact grains show size-dependent features that shift strongly with wavelength, whereas porous aggregates produce broader and qualitatively different spectra. 
Spatially resolved, multi-band full-Stokes observations with ALMA2040, analyzed jointly with the total intensity and spectral index, can therefore constrain the maximum aggregate size and filling factor. High angular resolution and broad wavelength coverage are essential for separating this signal from optical-depth effects and from polarization produced by aligned grains.

These measurements connect grain growth directly to planet formation. Together with the gas surface density, the inferred aggregate size and filling factor determine the Stokes number $\mathrm{St}$, the key dimensionless measure of dust--gas coupling. The Stokes number controls settling and radial drift. Together with the local solid abundance, it also sets the threshold for planetesimal formation via SI-driven clumping \citep{LiYoudin2021} and the pebble flux from which planetesimals feed from to grow to planet sizes \citep[i.e., pebble accretion;][]{Drazkowska2023ASPC..534..717D}. ALMA2040 can therefore test where disk solids are aerodynamically capable of forming planets. \\

\noindent {\bf D.~Tracing B-fields in disks} --  
Disk accretion is widely believed to be regulated by magnetohydrodynamic (MHD) turbulence driven by magnetorotational instability \citep[MRI;][]{Balbus1998} or by magnetically driven disk winds \citep{Blandford1982,Konigl2000,Lesur2021}.  Thus, B-fields are thought to play a direct role in planet formation. Detecting B-fields in protoplanetary disks has long been a primary science driver for ALMA, but clear characterization has remained elusive (Section~\ref{sec:understanding}).

ALMA2040 will substantially improve prospects for directly measuring B-fields in protoplanetary disks. Its multi-wavelength capabilities will identify optically thin dust emission, typically at longer wavelengths where disks are fainter, enabling separation of possible magnetic grain alignment from self-scattering. Its improved sensitivity will extend B-field studies to faint debris disks, where low gas densities reduce gas damping and allow grains to maintain alignment with the B-field. Despite marginal polarization detections via stacking toward $\beta$~Pic \citep{Hull2022}, debris disks remain largely unexplored. ALMA2040's improved linear and circular polarization sensitivity will enable the application of Comprehensive Zeeman Imaging (CZI), which combines both polarization signatures with Zeeman broadening to probe the 3-D morphology of disk B-fields \citep{LankhaarTeague2023}. The detection of Zeeman broadening in CN toward TW~Hya \citep{Teague2025} demonstrates the feasibility of this approach for planet-forming disks.




Similarly, methanol maser transitions at ALMA frequencies have been identified in massive protostellar disks \citep{Zinchenko2018,Sanna2021}. Although methanol is non-paramagnetic and consequently exhibits weak Zeeman splitting \citep{Lankhaar2018}, its extreme maser brightness makes it a sensitive probe of circular polarization. Reducing the systematic floor of Stokes~$V$ polarization measurements would therefore enable direct B-field measurements on disk scales.


\section{ALMA2040: Polarization Technical Requirements}
\subsection{Need for simultaneous, large-bandwidth observations for all ALMA frequencies}
Most studies require multi-band polarization observations to separate the scattering and thermal emission components (e.g., one band gives an incomplete picture; see Figures~1 and~2). A major leap in observing efficiency would be {\bf simultaneous observations of multiple bands, each with a large continuum bandwidth.} High spectral resolution ($<$1\,km\,s$^{-1}$) within these continuum observations would additionally allow a probe of linear and circular line polarization for free. The $\geq$512\,GHz instantaneous bandwidth with ALMA2040 would enable efficient multi-wavelength dust polarization and spectral line observations.
Compared to ALMA's current 7.5\,GHz, this large bandwidth would dramatically improve continuum sensitivity while simultaneously enabling high-precision, frequency-resolved measurements of how polarization varies across a band to accurately constrain scattering and the underlying emission mechanisms.

Currently, polarization observations are only available for Bands 1--7, despite ALMA technically having polarization capabilities at Bands 8-10.
Figure~\ref{fig:scattering_efficiency} shows that {\bf polarization is needed for ALMA's entire frequency range to properly model grain properties} for scattering. Moreover, since the GK effect requires sub-thermal lines and disks have midplane densities that are $>$10$^7$\,cm$^{-3}$, high critical density lines are needed. In Bands 8-10, abundant molecules such as HCN, HNC, and HCO$^+$ have critical densities of 10$^7$\,cm$^{-3}$ to a few 10$^{10}$\,cm$^{-3}$, 
making the GK effect more detectable in these bands. Bands 8--10 allow higher-frequency CO transitions, which are closer to sub-thermal, and atomic lines (e.g., [C~{\sc i}]) whose polarization probes B-fields via GSA.

\subsection{Need to Resolve Structure}
Polarization morphologies clearly change with substructure (see Figure~\ref{fig:pol_images}, right panel). As mentioned previously, the polarization morphology can also change with snow-lines, radial drift, and streaming instabilities, necessitating $\sim$sub-5\,au resolution for a large population of disks.

\begin{figure}[h!]
\includegraphics[width=0.99\textwidth,trim=2.1cm 16.4cm 0mm 4.2cm,clip=true]{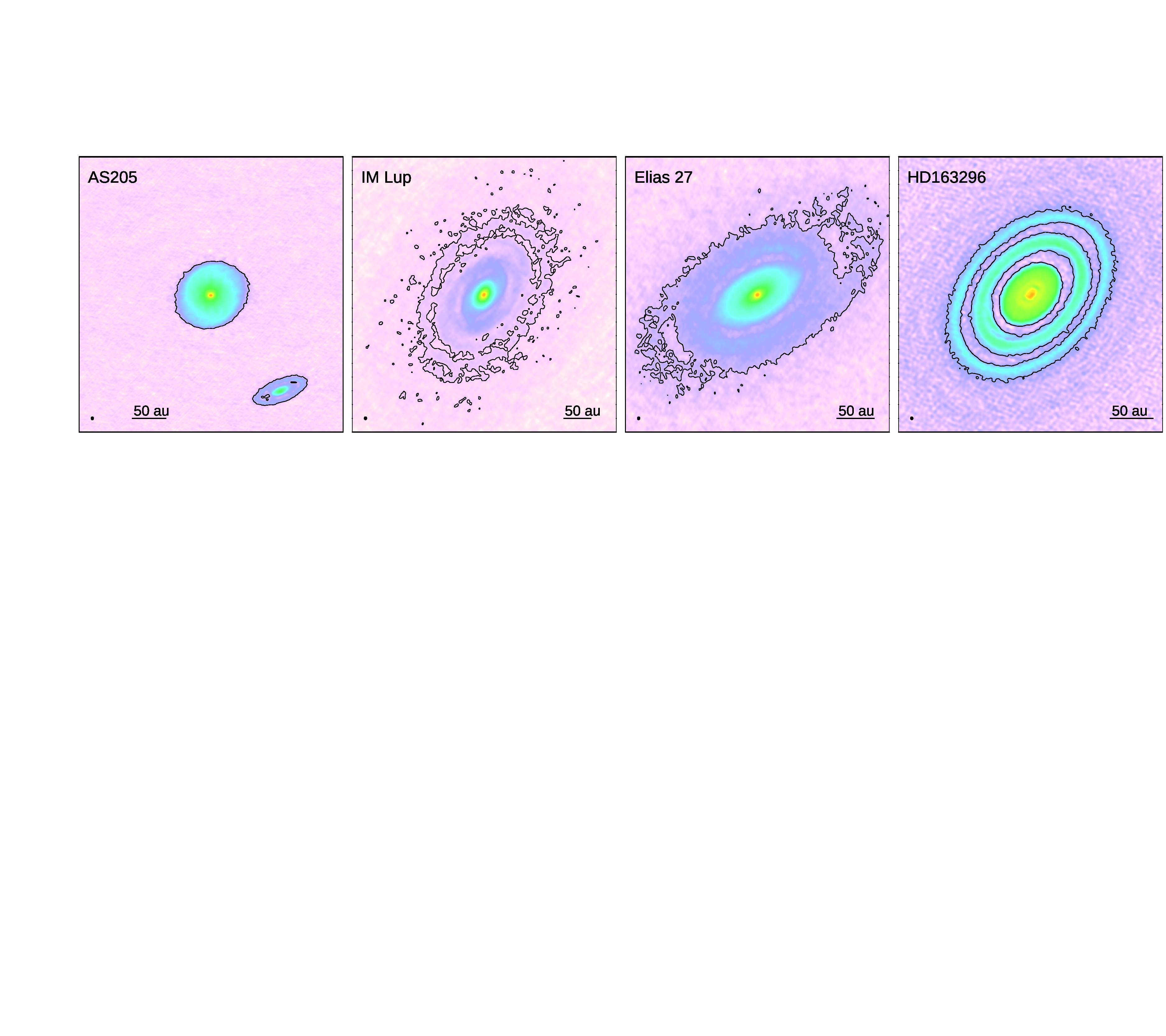}
\caption{Examples of a DSHARP-like dust polarization study at 239 GHz. Images show the current ALMA observations of these disks \citep{Andrews18} and contours show where 2\%\ dust polarization will be detected at $>$3\,$\sigma$ in 1.5 -- 7.8h on source with ALMA2040 assuming 16 GHz bandwidth. We simulated an ALMA2040 observation with \texttt{simobserve} using example intermediate and extended configurations to match a resolution of $\sim$5\,au. 
\label{fig:dsharp_2040}}
\end{figure}

The Disk Substructures at High Angular Resolution Project (DSHARP) survey targeted 20 nearby disks at $\sim$5\,au resolution at $\sim$240 GHz \citep{Andrews18}, revealing a breadth of substructure. ALMA2040 will enable an equivalent DSHARP survey in full dust polarization.  Figure~\ref{fig:dsharp_2040} shows the predicted polarization coverage for four DSHARP targets, with ALMA2040 simulated assuming a 12\,m array of 255 antennas that combines intermediate and extended configurations to match the 5\,au resolution of DSHARP. The contours indicate where 2\%\ polarization at $>$3$\sigma$ will be detected across the entire surface of the disk at 239 GHz, assuming a bandwidth of 16 GHz.  With only a few hours per source, the dust polarization can be resolved across each disk, including their rings and spiral arms. ALMA today would require a prohibitive $\sim$1.3\,$\times$\,$10^{4}$ hours on source for the entire DSHARP sample. \textbf{ALMA2040 could robustly detect dust polarization across all 20 of the DSHARP targets at $\sim$5\,au resolution in only 73 hours on source --- a factor of 180 improvement in integration time over ALMA today.}

Figure \ref{fig:dsharp_2040} shows a prediction assuming 16 GHz bandwidth centered at 239 GHz.  With $\geq$512\,GHz bandwidth, the aforementioned DSHARP-dust polarization study with ALMA2040 could simultaneously probe dust polarization at multiple wavelengths across each disk. 
This wide bandwidth also affords the flexibility to subdivide the data into narrower windows (e.g., 16 GHz) for targeted analysis since dust polarization varies with frequency  (see Figure \ref{fig:pol_images}). 


Eventually, observers will push ALMA2040 to its limits to probe substructure, particularly at $\sim$345\,GHz where disks are very bright. For an ALMA2040 345\,GHz sensitivity of 0.3\,$\mu$Jy\,bm$^{-1}$ for 24\,hr integration and 32\,GHz bandwidth, 1\% polarization can be detected at the 3$\sigma$ level in circumplanetary disks (CPDs) (peak $\sim$90\,$\mu$Jy\,bm$^{-1}$; \citealt{Isella2019}). Moreover, models of the polarized intensity of HL~Tau \citep{Stephens2023} suggest that this sensitivity will also allow 3$\sigma$ polarization to be detected across its entire disk for ALMA2040's anticipated maximum resolution (8.7 mas or 1.3 au). This resolution provides a factor of 16 improvement in solid angle over ALMA's current best \citep{Stephens2023} and would yield $\sim$20,000 independent measurements.



\subsection{Need for survey science}

Currently, we lack large enough samples to study the mechanisms behind dust polarization as a function of different environments, different disk evolutionary stages, and different initial conditions.  Only a few studies have observed multiple disks in dust polarization \citep[e.g.,][]{Lee2018,Sadavoy2019,Harrison19}, but these have been restricted to small samples of disks, moderate resolutions, and one or two wavelengths.

\begin{wrapfigure}[25]{r}{0.55\textwidth}
\vspace{-10pt}
\includegraphics[width=1\textwidth,trim=2.1cm 12.7cm 2cm 2.9cm,clip=true]{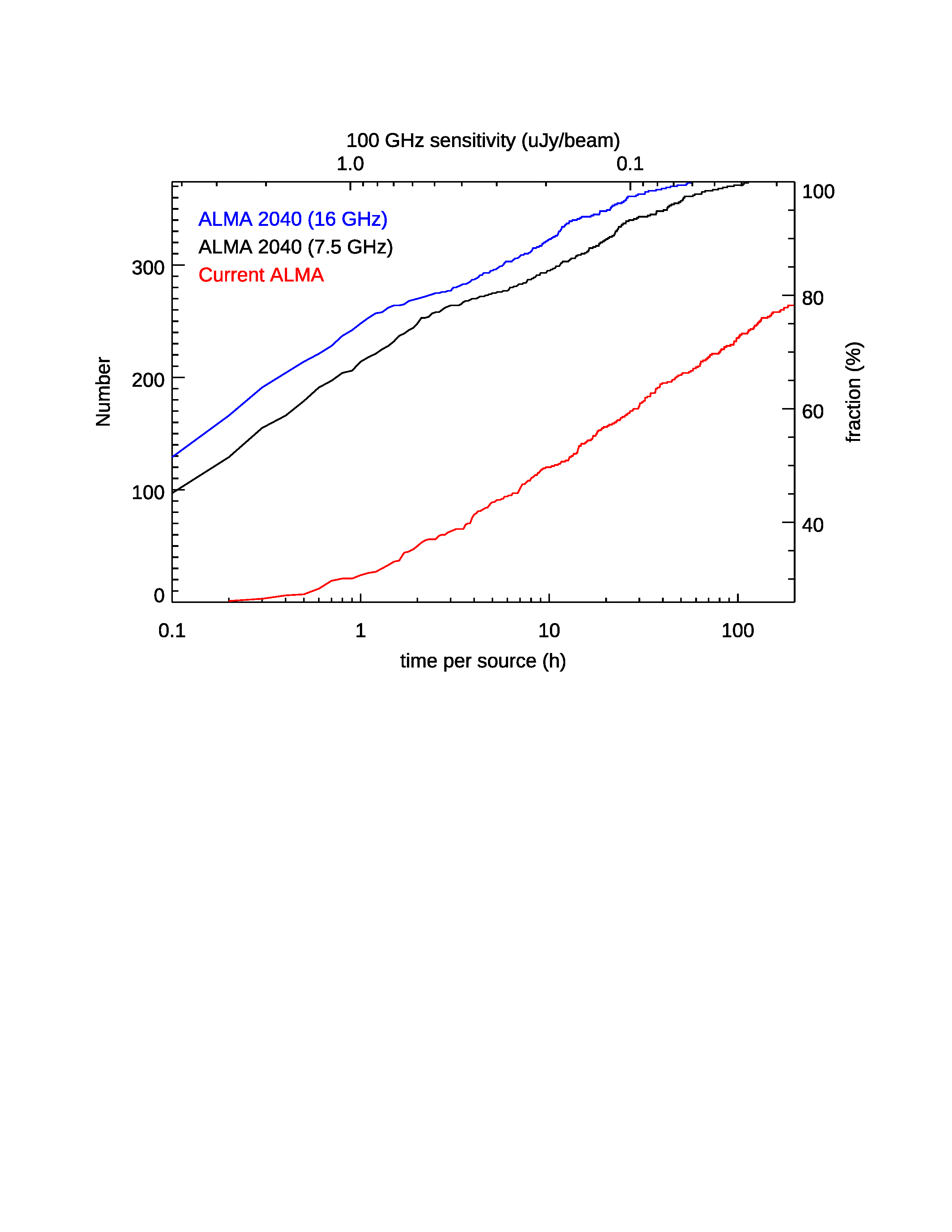}
\vspace{-20pt}
\caption{Comparison between ALMA2040 (blue and black) and current ALMA (red) to detect 1\% dust polarization at $>$3$\sigma$ at 100\,GHz toward 374 embedded disks in Orion.  We extrapolated the 370 GHz data from \citet{Tobin20} to 100\,GHz assuming $\alpha = 2$ and modelled the sensitivities using \texttt{simobserve} with the C8 (ALMA now) and an example intermediate ALMA2040 configuration to match 0.1\arcsec\ resolution. The curves show the number of disks that could be detected in polarization at 50\% of peak intensity.  
\label{fig:vandam_2040}}
\end{wrapfigure}

\textbf{With ALMA2040's improved sensitivity, large multi-wavelength surveys in dust polarization will be possible.} Figure \ref{fig:vandam_2040} compares the number of disks in Orion that can be well-detected at a 1\%\ polarization fraction with ALMA today (red) with what can be achieved with ALMA2040

\noindent (blue and black) at 100\,GHz (current Band~3) for a given time on source.  We used the ALMA Orion disk survey results from \citet{Tobin20}, which observed 374 disks at Band 7 and 0.1\arcsec\ resolution. Assuming 0.5 hour per source, we could robustly detect 1\% dust polarized emission at 0.1\arcsec\ resolution toward only 7 disks with ALMA's current sensitivity limits, whereas ALMA2040 would allow for $\sim$180 disks (48\% of the sample).  Even with 0.1 hour per disk, ALMA2040 will be able to observe almost 100 disks. ALMA2040's wide bandwidth would allow for polarization to be detected for  100s of disks in individual clouds at multiple bands, enabling multi-wavelength survey science for the first time.  


%
Large samples of disks can confirm whether or not dust scattering dominates at high frequencies or if such conditions are biased toward the densest disks \citep{Sadavoy2019}. Moreover, they can test the properties of dust grains, including sizes, porosity, and shape \citep[e.g.,][]{Zhang2023,Lin2025}, across different masses, environments, and evolutionary stages. Combining detailed studies of select disks and moderate resolution surveys of larger disk populations, ALMA2040 will probe how dust physics evolves over a disk lifetime and how dust growth and transport influence disk evolution.

\subsection{Need for spectral line polarization sensitivity}
Spectral lines may be the most promising avenue for detecting B-fields in planet-forming disks. Existing searches for Zeeman-induced Stokes~$V$ in CN have constrained line-of-sight B-fields to a few mG \citep{Vlemmings2019,Harrison2021}. Progress is limited not only by sensitivity, but also by polarimetric calibration: the current ALMA specification gives a minimum detectable circular polarization of $1.8\%$ of the peak Stokes~$I$. With the circular polarization scaling approximately as $P_V \propto B_{\rm los}\nu^{-1}$, and the absolute Stokes~$V$ scaling as $V\propto\nu$, high frequency Zeeman observations demand improved circular-polarization calibration, whereas low-frequency observations benefit from improved line sensitivity. 
The Zeeman effect provides more information than Stokes~$V$ alone. Zeeman broadening and Zeeman-induced linear polarization are second-order signatures, $\frac{\Delta v_Z}{\Delta v}\,,\ P_{L,Z}\propto B^2\nu^{-2}$,
and are therefore particularly powerful at the frequencies of Bands~1--3 \citep{LankhaarTeague2023}, where sufficient sensitivity would bring these signatures within observational reach. With ALMA2040, CZI can combine these signatures to measure the three-dimensional field strength and morphology of disk B-fields. 

For weakly accreting disks ($\dot{M}\sim10^{-9}$--$10^{-8}\,M_\odot\,\mathrm{yr}^{-1}$), MRI- and wind-driven accretion models require azimuthal fields of $\sim$3\,mG, and vertical magnetic fields of $\sim$0.3\,mG at radii of tens of au \citep{Bai2009,Lesur2021}. These levels remain beyond current Zeeman searches, which are limited by sensitivity \citep{Vlemmings2019} and by insufficient angular resolution to separate magnetic structures and mitigate beam smearing \citep{Harrison2021}. A tenfold improvement in line sensitivity would lower the TW~Hya limit from $ 0.8\,\mathrm{mG}$ to $\simeq0.08\,\mathrm{mG}$. Improvements in angular resolution would be equally important, 
enabling extraction of radial, toroidal, and vertical field components. With these components measured, 
we can directly constrain the magnetic stresses that drive accretion and 
differentiate between MRI- and wind-driven accretion models.

Improved spectral-line polarization sensitivity also enables B-field mapping through the GK effect and through collisional polarization. GK polarization requires sub-thermal excitation, making high-dipole-moment species such as HCN and the more highly excited CO transitions particularly promising disk tracers \citep{Lankhaar2022}. Collisional polarization predicts $\sim$0.5\% levels of polarization in molecular ion lines, for typical disk drift velocities \citep{Lankhaar2020}. The sensitivity gains envisioned for ALMA2040 would bring spatially resolved GK and collisional polarization within reach, providing B-field morphologies, as well as quantifying ambipolar diffusion, complementing Zeeman measurements.

Finally, extending spectral-line polarimetry to the higher frequencies (385--950\,GHz range) would provide access both to increasingly GK-sensitive, high-$J$ CO lines and to the atomic surface layers of planet-forming disks. There, ground-state atomic alignment (GSA) may produce B-field-dependent linear polarization in fine-structure lines such as [C~{\sc i}] \citep{Zhang2018}, which is also sensitive to Zeeman. A 10-fold increase in line sensitivity will allow [C~{\sc i}] linear and circular polarization to be detected at 3$\sigma$ levels of $<$1\% across entire disks such as IM Lup \citep{Law2023}.

\begin{wrapfigure}[16]{r}{0.4\textwidth}
    \centering
    \vspace{-55pt}
    \includegraphics[width=1\linewidth]{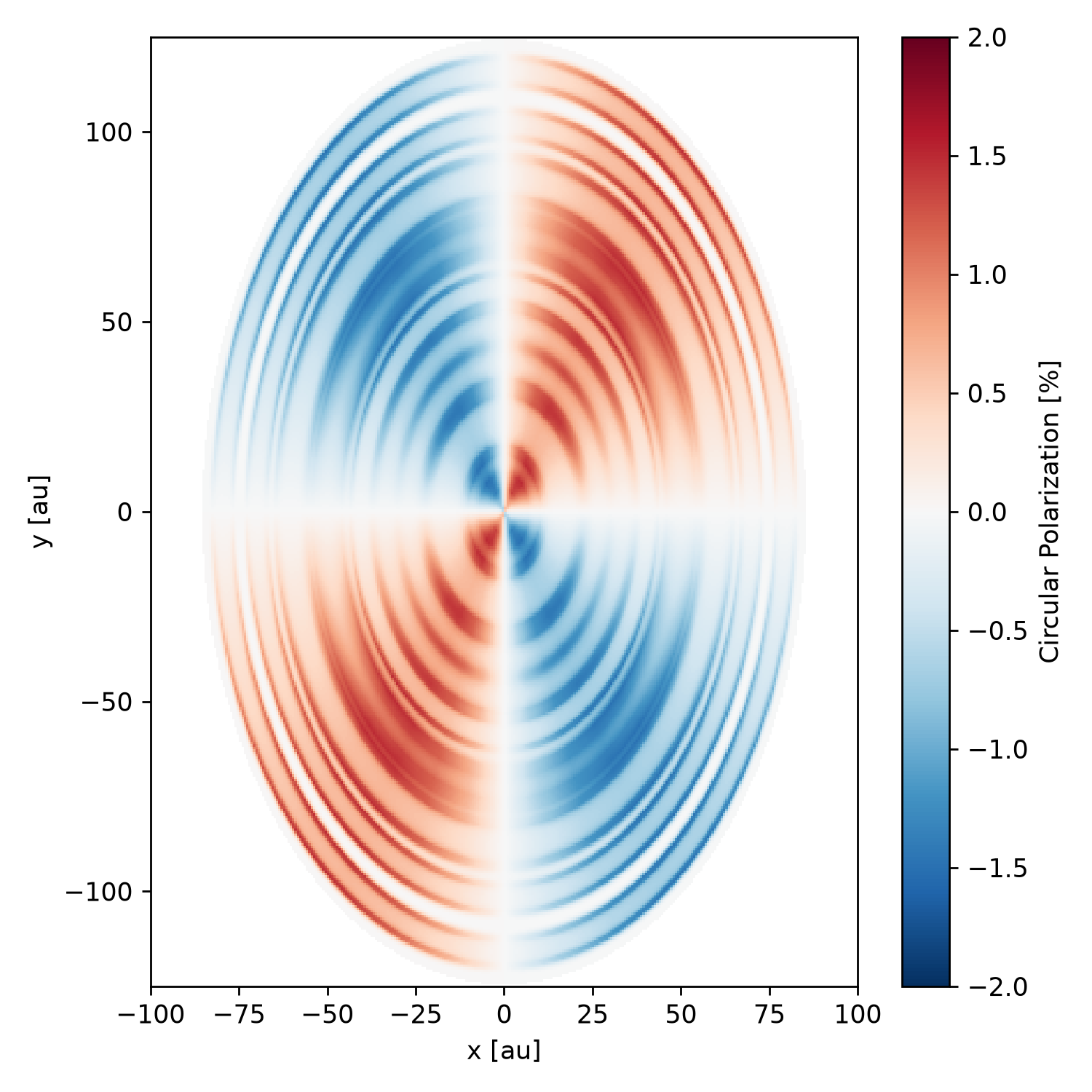}
    \vspace{-20pt}
    \caption{Predictions of circular polarization based on 
    the model of HL~Tau \citep{Stephens2023}. The magnitude depends on the optical depth, which alters between rings and gaps.}
    \label{fig:Stephens23_cpol}
\end{wrapfigure}

\subsection{Need for robust polarization calibration}
Polarization is critical to understanding planet formation and should be considered throughout all antenna instrumentation, rather than as an add-on. Well-characterized instrumentation will streamline calibration and reduce polarization commissioning time. 

\textbf{Full Polarization observations should become routine for ALMA for almost all observations.} Most disks are first observed deeply in total intensity, only to have a separate program to repeat the observations in Full Polarization. Making polarization a standard continuum data product would dramatically expand the sample of disks with polarimetric constraints at different wavelengths, where even upper limits would be valuable. {\bf By adding artificial polarization calibrators, the polarization calibration time can be reduced to a few minutes,} making polarization overheads negligible and eliminating the need for large parallactic angle coverage.


Current ALMA Stokes~$V$ observations are largely limited by beam-squint calibration limitations that restrict sensitivity to a 3$\sigma$ sensitivity of 1.8\% of the Stokes~$I$ flux. To make significant progress for Zeeman observations, the 3$\sigma$ level should be closer to 0.4\% \citep[e.g., to probe 1~mG fields;][]{Falgarone2008}, but ideally lower. In addition to B-fields, measurements of continuum Stokes~$V$ can further test predictions of grain physics. At (sub)millimeter wavelengths, circular polarization is expected to arise from scattering off aligned grains or from birefringence, each sensitive to the optical depth and grain properties such as size and shape. Figure~\ref{fig:Stephens23_cpol} shows the predicted circular polarization for HL Tau \citep{Stephens2023}, which is predominantly $\sim$0.1\% but can reach up to $\sim$1.5\% when the optical depth is of order unity. 
Moreover, beam squint and leakage should be well calibrated across the full primary beam to enable survey science and observations of disks with extended continuum and molecular-line emission.\\[-8mm]



\section{Synergies with current and future facilities}
Upcoming facilities operating at centimeter wavelengths, such as the ngVLA and SKAO, will provide a powerful complement to polarization studies in disks with ALMA2040, probing inner disk regions that may be optically thick for ALMA wavelengths, while better constraining disk grain properties (Figure~\ref{fig:scattering_efficiency}) and alignment mechanisms. At wavelengths longer than a few millimeters, additional mechanisms contribute to the continuum emission (including free-free, gyrosynchrotron, spinning dust) and provide valuable information on disk B-fields and winds \citep[e.g.,][]{Guidi2026SKA}. Combined, ALMA2040 and cm-wavelength facilities will separate these components, better constraining dust properties, polarization mechanisms, and disk magnetic fields. Moreover, complementary polarimetry in the infrared (e.g., ELT) would probe monomer-scale properties within dust aggregates.  In concert with ALMA2040, these facilities would provide a multi-scale view of dust properties in disks and dust aggregation leading to planet formation.\\[-8mm]

\section*{Acknowledgements}
This work was developed as part of the broader ALMA2040 community effort. We acknowledge the Lorentz Center and NOVA for supporting the community workshops and coordination activities that helped initiate and shape the ALMA2040 White Paper process. We acknowledge both referees for giving timely and helpful referee reports.\\

\bibliography{polrefs}
\bibliographystyle{apj_mod}

\end{document}